# The Physics and Applications of Superconducting Metamaterials

Steven M. Anlage[1,2]

[1]Center for Nanophysics and Advanced Materials, Physics Department, University of Maryland, College Park, MD 20742-4111

[2]Department of Electrical and Computer Engineering, University of Maryland, College Park, MD 20742

We summarize progress in the development and application of metamaterial structures utilizing superconducting elements. After a brief review of the salient features of superconductivity, the advantages of superconducting metamaterials over their normal metal counterparts are discussed. We then present the unique electromagnetic properties of superconductors and discuss their use in both proposed and demonstrated metamaterial structures. Finally we discuss novel applications enabled by superconducting metamaterials, and then mention a few possible directions for future research.

Our objective is to give a basic introduction to the emerging field of superconducting metamaterials. The discussion will focus on the RF, microwave, and low-THz frequency range, because only there can the unique properties of superconductors be utilized. Superconductors have a number of electromagnetic properties not shared by normal metals, and these properties can be exploited to make nearly ideal and novel metamaterial structures. In section I. we begin with a brief overview of the properties of superconductors that are of relevance to this discussion. In section II we consider some of the shortcomings of normal metal based metamaterials, and discuss how superconducting versions can have dramatically superior properties. This section also covers some of the limitations and disadvantages of superconducting metamaterials. Section III reviews theoretical and experimental results on a number of unique metamaterials, and discusses their properties. Section IV reviews novel applications of superconducting metamaterials, while section V includes a summary and speculates about future directions for these metamaterials.

## I. Superconductivity

Superconductivity is characterized by three hallmark properties, these being zero DC resistance, a fully diamagnetic Meissner effect, and macroscopic quantum phenomena.[1],[2],[3] The zero DC resistance hallmark was first discovered by Kamerlingh Onnes in 1911, and has since led to many important applications of superconductors in power transmission and energy storage. The second hallmark is a spontaneous and essentially complete diamagnetic response developed by superconductors in the presence of a static magnetic field. As the material enters the superconducting state it will develop currents to exclude magnetic field from its interior. This phenomenon is known as the Meissner effect, and distinguishes superconductors from perfect conductors (which would not show a spontaneous Meissner effect). Finally, macroscopic quantum effects arise from the quantum mechanical nature of the superconducting correlated electron state.

In some circumstances, the superconducting electrons can be described in terms of a macroscopic quantum wavefunction.[4] This wavefunction has a magnitude whose square is interpreted as the local density of superconducting electrons ($n_s$), and whose phase is coherent over macroscopic dimensions. This phase coherent wavefunction gives rise to quantum interference and tunneling effects which are quite extraordinary and unique to the superconducting state. Examples of macroscopic quantum phenomena include fluxoid quantization, and the DC and AC Josephson effects at tunnel barriers and weak links.

Superconductors have interesting and unique electromagnetic properties. Consider Fig. 1, which shows an idealized sketch of the complex conductivity of (a) a normal metal and (b) a superconductor. One can think of the electromagnetic response of a superconductor in terms of a crude but somewhat effective model, known as the two fluid model.[4] The superconductor consists of a super-fluid of Cooper-paired electrons coexisting with a second fluid of 'normal' or non-superconducting electrons. Electromagnetic fields interact with these two fluids in different ways. The superfluid has a purely reactive (inductive) response to external fields, and is responsible for establishing and maintaining the diamagnetic Meissner state below the critical temperature $T_c$. The normal fluid acts like a collection of ordinary electrons and is responsible for the bulk of the dissipative properties of the superconducting state, through absorption of photons.

Figure 1(a) shows a sketch of the normal-state conductivity, based on the Drude model in the Hagen-Rubens limit ($\omega\tau \ll 1$). The real part of the conductivity ($\sigma_1(f)$) starts at its DC value at zero frequency and monotonically decreases with increasing frequency, particularly at frequencies beyond the electron scattering rate $1/\tau$, well outside the range of frequencies considered here. The imaginary part of the normal state conductivity ($\sigma_2(f)$) is much smaller than the real part (by a factor of $\omega\tau \ll 1$), starts at zero at zero frequency, and shows a peak near $\omega \sim 1/\tau$. The small value of $\sigma_2$ translates into a near-zero value of the kinetic inductance of a normal metal, at least for frequencies $\omega \ll 1/\tau$. Hence the inductance of a normal metal wire is dominated by its geometrical inductance, which arises from energy stored in the magnetic field created by a normal current.

Figure 1(b) shows the conductivity of the same metal in the superconducting state in the local limit at zero temperature. The real part of the conductivity is now zero for frequencies between DC and the energy gap $2\Delta$, which is a measure of the binding energy of the Cooper pairs making up the superfluid. At zero frequency there is now a delta-function in $\sigma_1$, whose strength is proportional to the superfluid density in the material, $n_s$. This feature is responsible for the infinite conductivity at DC. Surprisingly, an ideal superconductor has zero AC (real) conductivity at zero temperature, between DC and the gap frequency. Losses arise from absorption of photons with energy greater than $2\Delta$, or from absorption by quasi-particles (broken Cooper pairs) that exist at finite temperature above the energy gap, or due to pair-breaking impurities in the material. The AC electrodynamics of a superconductor is dominated by the imaginary part of the conductivity, which at finite temperature is much greater than the real part in magnitude and is strongly frequency dependent ($\sigma_2 \sim 1/\omega$). The response of a superconductor is thus primarily diamagnetic and inductive. The inductance arises from screening currents that flow to maintain the Meissner state in the bulk of the material. These currents flow within a 'penetration depth' $\lambda$ of the surface. The penetration depth varies inversely with the superfluid density in the two-fluid model as $\lambda^2 = m/(\mu_0 n_s e^2)$, where $m$ and $e$ are the electronic mass and charge respectively, and $\lambda$ has values ranging from roughly 10's to 100's of nano-meters at zero temperature, depending on the material.

The superfluid density $n_s(T)$ is a monotonically decreasing function of temperature up to the critical temperature, where it is equal to zero, as such it acts as an order parameter of the superconducting state. Thus the penetration depth diverges as $T_c$ is approached from below, reflecting the weakening of the superconducting state and the penetration of magnetic flux deeper into the superconductor. The inductance of a superconductor has two contributions, one from the energy stored in magnetic fields outside and inside the superconductor, and the other from kinetic energy stored in the dissipation-less supercurrent flow. The latter contribution is known as kinetic inductance.[4] In the case of a thin and narrow wire with cross-sectional dimensions each less than $\lambda$, the kinetic inductance can be written approximately as:[5] $L_{Kin} = \mu_0 \lambda^2 / t$, where $t$ is the film thickness. As the superfluid density is diminished to zero (by means of increased temperature to $T_c$, or applied current to the critical current, or applied magnetic field to the critical field), the kinetic inductance will diverge. In this limit, when the superconductor is part of an electrical circuit it can make a dramatic change to the operation of that circuit as the kinetic inductance varies.

## I.1 Macroscopic Quantum Effects

As mentioned above, under certain circumstances a superconductor can be described by a macroscopic phase-coherent complex quantum wavefunction $\Psi = \sqrt{n_s} e^{i\theta}$. This wavefunction inherits its phase coherence from the underlying microscopic BCS wavefunction describing the Cooper pairing of all electrons in the metal. The absolute square of the macroscopic quantum wavefunction is interpreted as the local superfluid density[4] $n_s \sim |\Psi|^2$. When two superconductors are brought close together and separated by a thin insulating barrier, there can be quantum mechanical tunneling of Cooper pairs between the two materials.[1] This tunneling results in two types of Josephson effect. The first produces a DC current between the two superconductors which depends on the gauge-invariant phase difference between their macroscopic quantum wavefunctions as $I = I_c \sin(\varphi(t))$, where $\varphi(t) = \theta_1(t) - \theta_2(t) - \frac{2\pi}{\Phi_0} \int_1^2 \vec{A}(\vec{r},t) \bullet d\vec{l}$ is the gauge-invariant phase difference between superconductors 1 and 2, $\vec{A}(\vec{r},t)$ is the vector potential in the region between the superconductors, arising from magnetic fields, $I_c$ is the critical current of the junction, and $\Phi_0 = h/2e$ is the flux quantum ($h$ is Planck's constant and $e$ is the electronic charge). The DC Josephson effect states that a DC current will flow between two superconductors connected by a tunnel barrier, in the absence of a DC voltage, and the magnitude and direction of that current depend on a nonlinear function of the phase differences of their macroscopic quantum wavefunctions, as well as applied magnetic field.

The AC Josephson effect relates a DC voltage drop on the junction to a time-varying gauge-invariant phase difference, and therefore to an AC current in the junction: $d\varphi/dt = (2\pi/\Phi_0)V$, where $V$ is the DC voltage across the tunnel barrier. The AC impedance of a Josephson junction contains both resistive and reactive components in general.[6] At low frequencies and currents, the impedance of a Josephson junction can be treated with the resistively and capacitively shunted junction (RCSJ) model.[4] In the presence of both DC and AC applied currents in the junction, the nonlinear inductance of the junction can be described approximately as[4] $L_{JJ} = \Phi_0 / \left(2\pi I_c \sqrt{1 - \left(I_{DC}/I_c\right)^2}\right)$, where $I_{DC}$ is the DC current through the junction. This expression holds in the limit of small AC currents $I_{AC} << I_{DC}$.

By creating two Josephson junctions in parallel, forming a superconducting loop, it is possible to create quantum interference effects for electrical transport through the two junctions. Such a device is called a Superconducting Quantum Interference Device, or SQUID for short.[4] The critical current of a SQUID is a nonlinear function of applied magnetic flux in the SQUID loop, and its AC inductance is also a nonlinear function of applied field and current.[7]

Superconductors have two important length scales, the coherence length $\xi$, and the magnetic penetration depth $\lambda$, mentioned above. The coherence length is a measure of the characteristic scale for variations in the magnitude of the macroscopic quantum wave function. Like the penetration depth, it is a microscopic quantity at low temperatures and diverges as the critical temperature is approached from below. The ratio of these two length scales, the Ginzburg-Landau parameter $\kappa \equiv \lambda / \xi$, is approximately temperature independent, and distinguishes type-I superconductors ($\kappa < 1/\sqrt{2}$) from type-II superconductors ($\kappa > 1/\sqrt{2}$).[1] The difference between these two types of superconductors lies in the free energy required to create an interface between normal and superconducting phases inside the material. Type-II superconductors have a negative free energy for creating such interfaces, and thus the interfaces proliferate spontaneously when the material is exposed to a sufficiently strong magnetic field. The proliferation ends at the quantum limit, where each bounded normal phase unit supports exactly one quantum of magnetic flux, $\Phi_0$.

These quantized tubes of magnetic flux are known as magnetic vortices. They have a dynamics and interactions in superconductors that is quite interesting and non-trivial.[1,2] The electrodynamics of magnetic vortices stimulated by microwave currents has been studied extensively.[1,4,8] Vortices, when they are present, contribute both resistive and reactive contributions to the impedance of a superconductor.

Superconductivity can be found throughout the periodic table and in many compounds. Traditional (pre-1986) superconductors are generally referred to as "low-$T_c$" superconductors because their transition temperatures were in the range between a few mK to about 25 K. The "high-$T_c$" superconductors are mainly copper-oxide based, and have transition temperatures up to 150 K. More recently a number of interesting novel superconductors have been discovered in intermediate temperature ranges from 5 K to 50 K, including $MgB_2$,[9] and iron-pnictide superconductors.[10] Technologically important superconductors include Nb ($T_c$ = 9.2 K) and various alloys (e.g. Nb-Ti) that are used to create superconducting magnets for magnetic resonance imaging and particle accelerators.

Superconductivity exists within a dome in a three-dimensional parameter space spanned by temperature, electrical current density and magnetic field. The limit in temperature, $T_c$, is the temperature beyond which the superconducting electronic correlations are destroyed by thermal agitation. The limit in current is called the critical current density, $J_c$. This is a measure of when the kinetic energy of a super-current flow equals the free energy difference between the normal and superconducting states. The limit in magnetic field, the upper critical field ($H_{c2}$ for type-II superconductors) is a measure of when the magnetic vortices crowd close enough together such that their cores (of dimension $\xi$) begin to overlap and superconductivity is fully destroyed. In addition, there is a limit in the frequency domain beyond which superconducting electrodynamics alone cannot describe the surface impedance or complex conductivity of a superconductor. The gap frequency $f_{gap} = 2\Delta / h$ , is set by the 'energy gap' $2\Delta$ required to directly destroy a Cooper pair and create two quasi-particles. For most superconductors the gap frequency lies between about 10 GHz (low-$T_c$) and a few THz (high-$T_c$) at zero temperature. Note that the energy gap monotonically decreases with increasing temperature, going to zero at $T_c$.

## II. The Advantages of Superconducting Metamaterials

Metamaterials are typically constructed of "atoms" that have engineered electromagnetic response. The properties of the artificial atoms are often engineered to produce non-trivial values for the effective permittivity and effective permeability of a lattice of identical atoms. Such values include relative permittivities and permeabilities that are less than 1, close to zero, or negative. For concreteness, we shall consider below the scaling properties of metamaterials made of traditional "atomic" structures, like those used in the early metamaterials literature. Traditional metamaterials utilize wires to influence the dielectric properties by manipulating the effective plasma frequency of the medium.[11] The magnetic properties of Split-Ring Resonators (SRRs) are utilized to create a frequency band of sub-unity, negative or near-zero magnetic permeability.[12]

Substantial losses are one the key limitations of conventional metamaterials. As discussed in detail below, Ohmic losses place a strict limit on the performance of metamaterials in the RF – THz frequency range. In contrast to normal metals, superconducting wires and SRRs can be substantially miniaturized while still maintaining their low-loss properties.[13] For comparison, as the size of normal metal wires and SRRs are decreased, losses increase as $\rho/r^2$ and $\rho/t\ell$ , respectively, where $\rho$ is the resistivity of the metal, $r$ is the wire radius, $\ell$ is the size of the SRR, and $t$ is the thickness of the material making up the SRR.[14] Decreasing the normal metal wire radius of an $\varepsilon_{eff}(f) < 0$ artificial dielectric to increase its plasma frequency will result in significant increase of losses. As the normal metal SRR dimensions ( $t$ , $\ell$ ) decrease, losses will increase, and the frequency bandwidth of $\mu_{eff}(f) < 0$ will eventually vanish. These deleterious effects do not happen with superconducting wires and SRRs because the resistivity is orders of magnitude smaller, and the electromagnetic response is dominated by the reactive impedance. Superconductors will only break down when the dimensions become comparable to the coherence length, or when the induced currents approach the critical current density ($J_c \sim 10^6 - 10^9$ A/cm$^2$).

### II.1 Low Loss Metamaterials

Novel applications of negative index of refraction (NIR) metamaterials require that they be made very small compared to the wavelength and with very low losses. For example, the proposal of Ziolkowski and Kipple[15] to improve the radiation efficiency of ultra-small dipole antennas requires metamaterials with overall dimensions on the order of 1 mm operating at 10 GHz. Such metamaterials must be made with elements (wires and split-ring resonators) on the scale of 10's of micro-meters. Large losses will destroy the desired NIR behavior so care must be taken to see if existing metamaterials designs can be scaled down to the required dimensions. Here we examine such a scaling of a conventional normal metal metamaterial made up of split-ring resonators (to provide negative permeability) and wires (to provide negative permittivity).

### II.1.a Scaling of SRR Properties

Consider what happens when normal metal components are reduced in size to achieve the metamaterial dimensions mentioned above. Currently split-ring-resonators (SRRs) for operation at 10 GHz are fabricated using very thick plated metal (thickness $t \sim 25\text{-}50$ μm) on lossy dielectric substrates (such as FR4 and G10). These SRRs (Fig. 2) have outer diameters of about $w = 2.5$ mm, and gap widths on the scale of $g = 300\text{-}500$ μm.[16]

Let us examine what happens if the size of the SRR is reduced by a factor of 10, or more. The effective relative magnetic permeability of an SRR array is given by[12]

$$\mu_{eff} = 1 - \frac{\pi r^2 / \ell^2}{1 + i\frac{2\ell\rho_1}{\omega r\mu_0} - \frac{3\ell c_0^2}{\pi\omega^2 r^3 \ln\frac{2c}{d}}}, (1)$$

where $r$ is the inner radius of the inner SRR, $\ell$ is the lattice spacing of the cubic SRR array, $c_0$ is the speed of light, and $c$ and $d$ are defined in Fig. 2.

First, the resonant frequency of the SRR is given by; $\omega_0 = c_0\sqrt{\dfrac{3\ell}{\pi r^3 \ln\dfrac{2c}{d}}}$. The region of Re[ $\mu_{eff}$ ] < 0 is just above this resonant frequency. This suggests that if the lattice parameter scales with the radius $\ell \sim r$ and the ratio $c/d$ is kept fixed, the resonant frequency of the SRR will scale roughly as $\omega_0 \sim 1/r$ as $r$ decreases. However, this expression lacks the capacitive effect of the gap $g$, which will modify the scaling dependence. One can make $g$ quite small to increase the capacitance and keep the SRR resonant frequency at 10 GHz as the SRR shrinks.

Equation (1) shows the losses in the metal ring are parameterized by the sheet resistance per unit width $\rho_l = \rho/(ct)$, where $\rho$ is the resistivity of the metal and $t$ is the film thickness. The loss term (imaginary part) in $\mu_{eff}$ Eq. (1) is proportional to $\ell\rho_1/r \sim \ell\rho/(crt)$. Since the radius of the SRR ($r$) will scale like the lattice parameter $\ell$ as the SRR array is shrunk, the loss will scale roughly as $\rho/ct$. As the size of the SRR decreases, the width of the SRR $c$, and thickness of the film $t$ must also diminish. The thickness must decrease because one will not be able to make SRRs with dimensions c, d, g $\sim$ 5 μm using existing films with thickness t $\sim$ 50 μm. As the film thickness and SRR dimension shrink, the loss term in $\mu_{eff}$ will grow dramatically. For example, an order of magnitude decrease in SRR dimension (from $w = 2.5$ mm to 250 μm) will produce roughly a factor of 50-100 increase in loss. The situation is worse because the losses are dominated by the skin effect in the GHz range and beyond. In this limit the losses are proportional to the surface resistance ( $R_s = \rho/\delta$ in the local limit, where $\delta$ is the skin depth), and the skin depth is typically smaller than the film width and thickness. To achieve the objectives of the proposed antenna application, even further shrinkage of the SRR would be required. Normal metal losses are already significant in conventional designs, so clearly normal metal SRRs will not be practical for ultra-small metamaterials.

### II.1.b Scaling of Wire Array Properties

A wire array of lattice spacing $a$ and wire radius $r$ has a plasma frequency given approximately by,[11] $\omega_p = \dfrac{c_0\sqrt{2\pi}}{a\sqrt{\ln(a/r)}}$. The real part of the effective permittivity of the wire array is negative for frequencies below the plasma frequency. Again, as the wire array lattice parameter shrinks, the plasma edge will increase roughly as $\omega_p \sim 1/a$. This scaling will maintain an $\varepsilon_{eff} < 0$ range at 10 GHz, as desired. The complex effective relative electric permittivity is given by;[11] $\varepsilon_{\text{eff}} = 1 - \dfrac{\omega_p^2}{\omega\left(\omega + i\varepsilon_0 a^2 \omega_p^2 / \pi r^2 \sigma\right)}$. The loss term is given by $\varepsilon_0 a^2 \omega_p^2 \rho / \pi r^2$, where $\rho = 1/\sigma$ is the resistivity of the normal metal wire. Given the fact that $\omega_p \sim 1/a$, this leads to a dielectric loss that scales as $\rho / r^2$ as the wire radius shrinks. Again one runs in to trouble as the metamaterial is shrunk, with a strong inverse-square increase of losses as the wire radius is diminished.

It is clear that the properties of conventional normal metal SRR and wire arrays suffer from a dramatic increase in loss as their dimensions are uniformly shrunk to achieve sizes required for non-trivial applications. This makes normal metals impractical for any application involving an order of magnitude or more decrease in the size of a metamaterial that begins with minimally acceptable losses. Alternatives to this dilemma must be sought, and superconducting metamaterials are one obvious choice.

## II.2 Compact Superconducting Metamaterials

Superconductors have small values of the surface resistance at microwave frequencies.[2] Typical surface resistance values for Nb at 2 K and 1 GHz are in the nano-Ohm range,[17] while high temperature superconductors have a surface resistance on the scale of 100 micro-Ohms at 77 K and 10 GHz.[18] These small losses (and associated $\sigma_2$ -dominated electrodynamics) allow superconductors to overcome the scaling limitations discussed above. In addition, superconductors have other unique advantages, and these advantages are especially evident when they are scaled down in size. When the cross-sectional area of a current-carrying superconductor is reduced such that one or both are on the scale of the magnetic penetration depth λ, the kinetic inductance will be enhanced. As discussed above and below, this is the inductance associated with the kinetic energy of the dissipation-less supercurrent flow. If the cross-sectional area is reduced so that the width or thickness is on the scale of the coherence length ξ, or if a narrow tunnel barrier is created in the superconductor, then the Josephson effect also comes in to play.[1] Thus a new contribution to the inductance can arise from currents flowing through a Josepshon junction, $L_{JJ}$. These two forms of inductance allow superconducting metamaterials to be even further miniaturized, compared to normal metal-based metamaterials.

## II.3 Tunability and Nonlinearity

Because of the unique quantum-mechanical properties of superconductors, their inductance is highly tunable and can be made quite nonlinear. First, the superfluid density can be a strong function of applied magnetic field (both DC and RF) and current, creating the opportunity to vary the kinetic and Josephson inductances. The use of temperature-dependent tuning of a negative permeability region of superconducting split ring resonators has been demonstrated.[13,14,19] The use of a DC current to tune the kinetic inductance of superconductors has also been considered.[5,20]

Superconductors also support quantized magnetic vortex excitations, and these bring with them inductance and loss when agitated by RF currents.[8] Subsequent work on superconducting metamaterials showed that the negative permeability region can also be tuned by changing the RF current circulating in the SRRs.[21] This was attributed to RF magnetic vortices entering into the SRRs at sharp inside corners of the patterned structure. This hypothesis was confirmed through imaging of the RF current flow in a high-temperature superconducting SRR.[21,22] DC magnetic vortices have also been added to superconducting SRRs, and frequency shifts in the negative permeability region have been observed.[21] It was hypothesized that the frequency shifts came about from an inhomogeneous distribution of magnetic flux entering and leaving the SRR as the field was ramped. Microscopic imaging of the flux penetration by means of magneto-optical imaging[23] verified the strongly inhomogeneous nature of flux penetration.[21] Magnetic vortex lattices in superconductors have also been proposed as a way to generate photonic crystals, where the dielectric contrast is created between the superconducting bulk and non-superconducting cores of the vortices.[24, 25] Such materials are considered in more detail below.

The Josephson inductance is a strong function of applied currents and fields. Tuning of Josephson inductance in a transmission line geometry was considered theoretically by Salehi.[20, 26] The tunability of SQUID inductance by external fields was considered by Du, *et al.*[27] The Josephson inductance is also known to be very sensitive to vortices propagating in an extended Josephson junction.[28, 29, 30]

## II.4 The Limitations of Superconducting Metamaterials

The greatest disadvantage of superconductors is the need to create and maintain a cryogenic environment, and to bring signals to and from the surrounding room temperature environment. The use of liquid cryogens (such as Nitrogen and Helium) is inconvenient and increasingly expensive. Closed-cycle cryocooler systems have become remarkably small, efficient and inexpensive since the discovery of high-$T_c$ superconductors. Such systems are now able to operate for 5 years un-attended, and can accommodate the heat load associated with microwave input and output transmission lines to room temperature.[31]

Superconductors can also be very sensitive to environmental influences. Variations in temperature, stray magnetic field, or strong RF power can alter the superconducting properties and change the behavior of the metamaterial. Careful temperature control and high quality magnetic shielding are often required for reliable performance of superconducting devices.

## III. Novel Superconducting Metamaterial Implementations

A number of novel implementations of superconducting metamaterials have been achieved in addition to superconducting split rings and wires,[13, 14, 21]. Here we present results on several classes of superconducting metamaterials.

## III.1 Superconductor/Ferromagnet Composites

Ferromagnetic resonance offers a natural opportunity to create a negative real part of the effective permeability of a gyromagnetic material for frequencies above resonance.[32] However the imaginary part of the permeability is quite large near the resonance and will limit the utility of such Re[$\mu_{eff}$] < 0 materials. Combining such a material with a superconductor can help to minimize losses,[33] and to introduce a conducting network with Re[$\varepsilon_{eff}$] < 0 at the same time. A superlattice film of high temperature superconductor and manganite ferromagnetic layers was created and shown to produce a band of negative index in the vicinity of 90 GHz.[34] The material displayed Re[n] < 0 near an applied field of 3 T at 90 GHz, although the imaginary part of the index (Im[n]) was of comparable magnitude to the real part.

## III.2 DC Magnetic Superconducting Metamaterials

The concept of a DC magnetic cloak has been proposed and examined by Wood and Pendry.[35] The general idea is to take a solid diamagnetic superconducting object and divide it into smaller units, arranging them in such a way as to tailor the magnetic response. The cloak would shield a region of space from external DC magnetic fields, and not disturb the magnetic field distribution outside of the cloaking structure. The cloak involves the use of superconducting plates to provide diamagnetic response, causing the radial component of effective permeability ($\mu_r$) to lie between 0 and 1. An additional component is required to enhance the tangential components of magnetic permeability so that $\mu_\theta$ and $\mu_\phi$ are both greater than 1, and paramagnetic substances were suggested in the original proposal. An experimental demonstration of the first step ($\mu_r < 1$) in creating such a cloak was made using an arrangement of Pb thin film plates.[36] Subsequent theoretical work has refined the DC magnetic cloak design and suggested that it be implemented with high-temperature superconducting thin films.[37]

### III.3 SQUID Metamaterials

A SQUID is a natural quantum analog of the split ring resonator. In fact the Radio Frequency (RF) SQUID, developed in the 1960's, is essentially a quantum SRR in which the classical capacitor is replaced with a Josephson junction. It's original purpose was to measure small RF magnetic fields and operate as a flux to frequency transducer.[38],[39] The first proposal to use an array of RF SQUIDs as a metamaterial was made by Du, Chen and Li.[40],[27] Their calculation assumes the SQUID has quantized energy levels and considers the interaction of individual microwave photons with the lowest lying states of the SQUID potential. For small detuning of the microwave photon frequency above the transition from the ground state to the first excited state, the medium will present a negative effective permeability. However, the frequency region of negative permeability is diminished by a non-zero de-phasing rate, and negative permeability will disappear for de-phasing rates larger than a critical value.

A treatment of RF SQUIDs interacting with classical electromagnetic waves was presented by Lazarides and Tsironis.[41] They considered a two-dimensional array of RF SQUIDs in which the Josephson junction was treated as a parallel combination of resistance, capacitance and Josephson inductance. Near resonance, a single RF SQUID can have a large diamagnetic response. In an array, there is a frequency and RF-magnetic field region in which the system displays a negative real part of effective permeability. The permeability is in fact oscillatory as a function of applied magnetic flux, and will be suppressed with applied fields that induce currents in the SQUID that exceed the critical current of the Josephson junction. Related work on a one-dimensional array of superconducting islands that can act as quantum bits (qubits) was considered by Rakhmanov, et al.[42] When interacting with classical electromagnetic radiation, the array can create a quantum photonic crystal that can support a variety of nonlinear wave excitations. A similar idea based on a SQUID transmission line was implemented to perform parametric amplification of microwave signals.[43],[44]

### III.4 Radio Frequency Superconducting Metamaterials

Superconducting thin film wires can create large inductance values without the associated losses found in normal metals. This offers the opportunity to extend metamaterial atom structures to lower frequencies where larger inductance values are required to build resonant structures. In the past, three-dimensional structures have been employed to create artificial magnetism below 100 MHz.[45] The use of two-dimensional spiral resonators to create negative effective permeability atoms has been developed in the sub-GHz domain using thick normal metal wires.[46] However, such spirals are too lossy to resonate below 100 MHz, for reasons similar to those discussed above in section II.1. Superconducting thin film spirals have low losses, but also have enhanced inductance from the kinetic inductance of the superfluid flow. Superconducting spiral metamaterials operating near 75 MHz have been developed with Nb thin films, and show strong tunability as the transition temperature is approached.[47] Such metamaterials may be useful for magnetic resonance imaging,[45,48] near-field imaging,[49] and compact RF resonator applications.[50]

### III.5    Superconducting Photonic Crystals

Photonic crystals (PCs) are generally constructed from a modulated dielectric function contrast with a spatial scale on the order of the wavelength.[51] As such, they fall outside the domain of what are usually called metamaterials, but we shall consider them here nonetheless. PCs show intricate band structure arising from multiple scattering of light from the dielectric contrast in the material. Superconducting photonic crystals have been proposed and discussed by a number of researchers[24,25,52,53,54,55,56,57,58,59,60,61] The original suggestions for superconducting PCs pre-dated most of the work on metamaterials. The earliest ideas proposed propagating modes in anisotropic cuprate superconducting PCs,[24] and a new spectroscopic feature (a phonon-polariton-like gap) in a superconducting PC.[25,52] In the latter case, it was found that the gap is temperature dependent, due to the temperature dependence of the superfluid density, and therefore tunable.[52,53,54,62,57]

High temperature cuprate superconducting materials have strongly anisotropic dielectric properties. They show metallic behavior for currents flowing in the Cu-O planes, and have dielectric-like properties for light polarized in the c-direction.

As such, they can be used for both PC structures[24,54] and metamaterials.[63] It is known that the c-axis plasma frequency can be tuned by a DC magnetic field,[64,65] hence the entire PC band structure can also be tuned by a magnetic field.[54]

Feng, et al.,[66] examined a PC made up of superconducting cylinders embedded in a dielectric medium. They found that temperature tuning of the superfluid density, and therefore of the effective superfluid plasma frequency, can produce a tunable band of all-angle negative refraction, as well as tunable refracting beams. The angle of refraction can be tuned from positive to negative values, and achieve up to 45° of sweep. This work did not consider the losses or plasmonic effects due to quasiparticles, and did not take into account the frequency bandwidth limitations imposed by the superconducting gap 2Δ. Subsequent work showed that neglect of the quasiparticles introduces significant errors, although a strongly temperature dependent quasiparticle scattering rate (as seen in certain cuprate superconductors[67]) can restore some of the temperature tunability.[58]

Development of dielectric contrast from an Abrikosov vortex lattice to create a photonic crystal is a novel idea first proposed by Takeda, et al.,[68] and studied by Kokabi, et al.,[69] and Zandi, et al.[70] Magnetic vortices form a long-range ordered lattice in an ideal superconductor, creating a regular dielectric contrast between normal metal core tubes and the superconducting background. Takeda, et al. assumed the cores were described by a constant permittivity, while outside it is a dissipation-less Drude metal. They found that the PC band structure, and gaps, can be tuned by variation of the magnetic field (which changes the PC lattice parameter) and the Ginzburg-Landau parameter, κ.[68] Kokabi, et al.,[69] and Zandi, et al.[70] calculated the superconducting electron density using the self-consistent Ginzburg-Landau formalism, and calculated the resulting PC band structure for both isotropic[69] and anisotropic[70] superconductors.

## IV.  Novel Applications Enabled by Superconducting Metamaterials

Superconducting metamaterials have been utilized to realize a number of unique applications.  In addition, theorists have made predictions for other exciting applications of superconducting metamaterials which have not yet been realized.

The combination of left-handed and right-handed propagation media creates opportunities for new types of resonant structures.  Engheta predicted that a resonant structure created by laminating two materials with opposite senses of phase winding can create a new class of resonant structures.[50]  He proposed a resonator consisting of two flat conducting plates separated by a sandwich of left-handed and right-handed metamaterials.  A wave propagating in the direction normal to the plates will suffer a combination of forward and reverse phase windings before reaching the other reflecting boundary.  Under these conditions the wave could undergo a net phase shift of 0 radians and still create a resonance condition.  The net phase shift could also be a positive or negative multiple of $2\pi$ radians as well, each creating a resonant condition.  The result is an ultra-compact resonator whose overall dimension is no longer constrained by the wavelength of the resonant wave.  The first realization of this ultra-compact resonator with superconductors was accomplished by Wang and Lancaster.[19,71]  They utilized a lumped-element dual transmission line structure implemented in a co-planar waveguide geometry.  Such structures have proven very useful for creating practical backward-wave microwave devices,[72] and superconducting versions were also studied by Salehi, Majedi and Mansour.[20]  The superconducting sub-wavelength resonator was implemented with cuprate superconducting films and showed resonances with indices between +2 and -6, including zero, over the broad frequency range from about 5 to 24 GHz.  Superconductors are particularly attractive for use in ultra-compact resonators because otherwise the Ohmic losses of a deep sub-wavelength structure would suppress the Q and render the device practically useless.  Quality factors of the negative order resonances were on the order of 3000 at 30 K, while those for positive order resonances were below 400.[19]

A related theoretical proposal was to add Josephson inductance to the superconducting dual transmission line to enhance nonlinearity. This can result in a tunable dispersion relation for propagating waves in the structure.[26]  Considering a metamaterial made up of a Josephson quantum bit (qubit) array leads to a number of interesting predictions, including the development of a quantum photonic crystal that derives its properties from the quantum states accessible to the qubits.[42]

A one-dimensional SQUID array nonlinear transmission line has been used as a parametric amplifier, tunable for microwave signals between 4 and 8 GHz, and providing up to 28 dB of gain.[43,44]  This amplifier is suitable for use in detecting signals from low temperature qubits operating at microwave frequencies, and can squeeze quantum noise.[44]

## V.  Future Directions and Conclusions

There are many exciting future directions of research in superconducting metamaterials.  Their low loss, compact structure, and nonlinear properties make them ideal candidates for realization of the landmark predictions of metamaterial theory, including the near-perfect lens, evanescent wave amplification, hyper-lensing, transformation optics and illusion optics.

There are many exciting possibilities to extend the frequency coverage of superconducting metamaterials from the low-RF range (below 10 MHz) to the upper limits of superconductivity in the multi-THz domain. Low frequencies offer the opportunity to create deep sub-wavelength structures to act as energy storage devices, imaging devices, or as filters. The THz domain brings many exciting possibilities for spectroscopy and imaging as well.

Josephson-based metamaterials offer many interesting opportunities for novel meta material structures. Their nonlinear response can be used to introduce parametric amplification[73] of negative-index photons, further reducing the deleterious effects of loss. They also have extreme tunability with both DC and RF magnetic fields due to changes in the Josephson inductance. Their properties can also be made broad-band using a plurality of junction critical currents in the design.

Many superconductors have intrinsic properties that make them quite suitable for use as metamaterials or photonic crystals. For example, the anisotropic dielectric function in the layered high-$T_c$ cuprate superconductors[63] offers an opportunity to realize a hyperlens in the THz and far-infrared domain.[74] The cuprates also have a built-in plasmon excitation for electric fields polarized along their c-axis (the nominally insulating direction),[64],[65] which can be exploited for plasmonic applications. Conventional superconductors have also shown conventional propagating plasmon excitations,[75] suggesting that low loss microwave plasmonics, analogous to optical plasmonics, can be developed using superconducting thin films.

**Acknowledgements**: This project has been supported by the US Office of Naval Research through grant # N000140811058, and the Center for Nanophysics and Advanced Materials (CNAM) at the University of Maryland.

Figure Captions

Fig. 1 Schematic plot of the complex conductivity of (a) normal metal and (b) superconductor versus frequency ω in the Hagen-Rubens limit (ωτ << 1). We assume ideal materials at zero temperature in this sketch.

Fig. 2 (a) A "single" planar split-ring resonator (SRR) with dimensions of the film (black) shown. Based on the sketch in Ref. ([16]). (b) A stack of SRRs illustrating the lattice spacing ℓ.

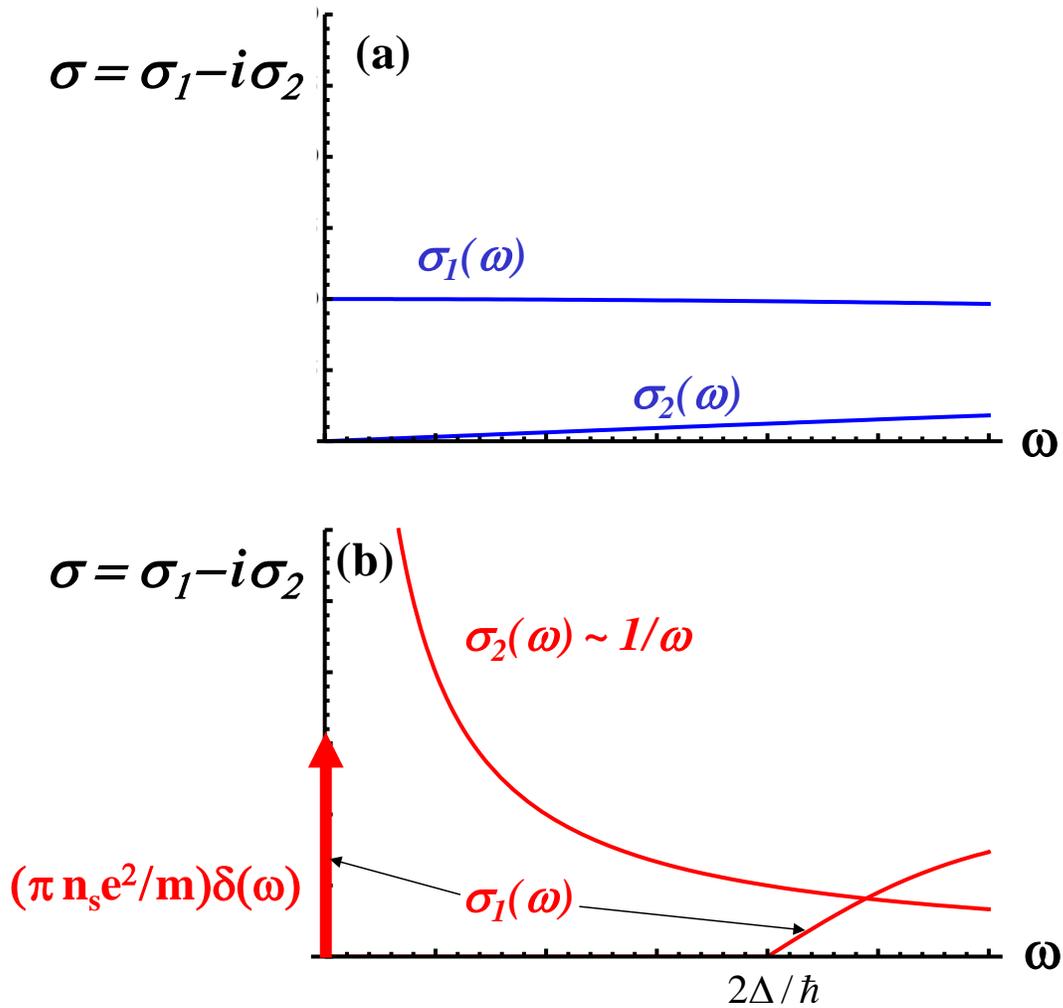

Figure 1

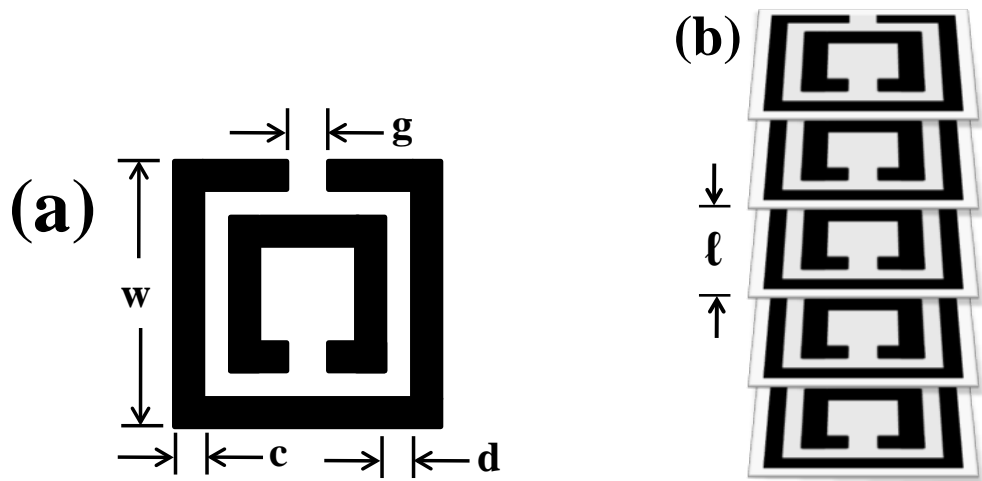

Figure 2

**References**:

1 M. Tinkham, *Introduction to Superconductivity*. (McGraw-Hill, New York, 1996), 2nd Edition ed.

2 Alan M. Portis, *Electrodynamics of High-Temperature Superconductors*. (World Scientific, Singapore, 1993).

3 J. B. Ketterson and N. Song, *Superconductivity*. (Cambridge University Press, Cambridge, 1999).

4 T. P. Orlando and K. A. Delin, *Foundations of Applied Superconductivity*. ( Addison-Wesley, Reading, MA, 1991).

5 S. M. Anlage, H. J. Snortland, and M. R. Beasley, "A Current Controlled Variable Delay Superconducting Transmission-Line," Ieee T Magn **25** (2), 1388-1391 (1989).

6 F. Auracher, P. L. Richards, and G. I. Rochlin, "Observability of Quasiparticle-Pair Interference Current in Superconducting Weak Links," Phys Rev B **8** (9), 4182-4185 (1973).

7 O. G. Vendik and I. S. Danilov, "Equivalent parameters of a Josephson junction in a microwave SQUID structure," Tech Phys+ **44** (11), 1360-1366 (1999).

8 M. W. Coffey and J. R. Clem, "Unified Theory of Effects of Vortex Pinning and Flux Creep Upon the Rf Surface Impedance of Type-Ii Superconductors," Phys Rev Lett **67** (3), 386-389 (1991).

9 C. Buzea and T. Yamashita, "Review of the superconducting properties of MgB2," Superconductor Science and Technology **14** (11), R115-R146 (2001).

10 K. Ishida, Y. Nakai, and H. Hosono, "To What Extent Iron-Pnictide New Superconductors Have Been Clarified: A Progress Report," J Phys Soc Jpn **78** (6), 062001 (2009).

11 J. B. Pendry, A. J. Holden, W. J. Stewart, and I. Youngs, "Extremely low frequency plasmons in metallic mesostructures," Phys Rev Lett **76** (25), 4773-4776 (1996).

12 J. B. Pendry, A. J. Holden, D. J. Robbins, and W. J. Stewart, "Magnetism from conductors and enhanced nonlinear phenomena," Ieee T Microw Theory **47** (11), 2075-2084 (1999).

13 M. Ricci, N. Orloff, and S. M. Anlage, "Superconducting metamaterials," Appl Phys Lett **87** (3), 034102 (2005).

14 M. C. Ricci and S. M. Anlage, "Single superconducting split-ring resonator electrodynamics," Appl Phys Lett **88** (26), 264102 (2006).

15 R. W. Ziolkowski and A. D. Kipple, "Application of double negative materials to increase the power radiated by electrically small antennas," Ieee T Antenn Propag **51** (10), 2626-2640 (2003).

16 R. A. Shelby, D. R. Smith, S. C. Nemat-Nasser, and S. Schultz, "Microwave transmission through a two-dimensional, isotropic, left-handed metamaterial," Appl Phys Lett **78** (4), 489-491 (2001).

17 J. Knobloch H. Padamsee, T. Hays, *RF Superconductivity for Accelerators*. (J. Wiley, New York, 1998).

18 V. V. Talanov, L. V. Mercaldo, S. M. Anlage, and J. H. Claassen, "Measurement of the absolute penetration depth and surface resistance of superconductors and normal metals with the variable spacing parallel plate resonator," Rev Sci Instrum **71** (5), 2136-2146 (2000).

19 Y. Wang and M. J. Lancaster, "High-temperature superconducting coplanar left-handed transmission lines and resonators," Ieee T Appl Supercon **16** (3), 1893-1897 (2006).

20 H. Salehi, A. H. Majedi, and R. R. Mansour, "Analysis and design of superconducting left-handed transmission lines," Ieee T Appl Supercon **15** (2), 996-999 (2005).

21 M. C. Ricci, H. Xu, R. Prozorov, A. P. Zhuravel, A. V. Ustinov, and S. M. Anlage, "Tunability of superconducting metamaterials," Ieee T Appl Supercon **17** (2), 918-921 (2007).

22 A. P. Zhuravel, A. G. Sivakov, O. G. Turutanov, A. N. Omelyanchouk, S. M. Anlage, A. Lukashenko, A. V. Ustinov, and D. Abraimov, "Laser scanning microscopy of HTS films and devices (Review Article)," Low Temp Phys+ **32** (6), 592-607 (2006).

23 C. Jooss, J. Albrecht, H. Kuhn, S. Leonhardt, and H. Kronmuller, "Magneto-optical studies of current distributions in high-T-c superconductors," Rep Prog Phys **65** (5), 651-788 (2002).

24 W. M. Lee, P. M. Hui, and D. Stroud, "Propagating photonic modes below the gap in a superconducting composite," Phys Rev B **51**, 8634 (1995).

25 C. H. R. Ooi and T. C. A. Yeung, "Polariton gap in a superconductor-dielectric superlattice," Phys Lett A **259** (5), 413-419 (1999).

26 H. Salehi, R. R. Mansour, and A. H. Majedi, "Nonlinear Josephson left-handed transmission lines," Iet Microw Antenna P **1** (1), 69-72 (2007).

27 C. G. Du, H. Y. Chen, and S. Q. Li, "Stable and bistable SQUID metamaterials," J Phys-Condens Mat **20** (34), 345220 (2008).

28  D. E. Oates and P. P. Nguyen, "Microwave power dependence of YBa[sub 2]Cu[sub 3]O[sub 7] thin-film Josephson edge junctions," Appl Phys Lett **68** (5), 705-707 (2001).

29  S. C. Lee and S. M. Anlage, "Spatially-resolved nonlinearity measurements of YBa2Cu3O7-delta bicrystal grain boundaries," Appl Phys Lett **82** (12), 1893-1895 (2003).

30  S. C. Lee, S. Y. Lee, and S. M. Anlage, "Microwave nonlinearities of an isolated long YBa2Cu3O7-delta bicrystal grain boundary," Phys Rev B **72** (2), 024527 (2005).

31  R. W. Simon, R. B. Hammond, S. J. Berkowitz, and B. A. Willemsen, "Superconducting microwave filter systems for cellular telephone base stations," P Ieee **92** (10), 1585-1596 (2004).

32  S. T. Chui and L. B. Hu, "Theoretical investigation on the possibility of preparing left-handed materials in metallic magnetic granular composites," Phys Rev B **65** (14), 144407 (2002).

33  T. Nurgaliev, "Modeling of the microwave characteristics of layered superconductor/ferromagnetic structures," Physica C: Superconductivity **468** (11-12), 912-919 (2008).

34  A. Pimenov, A. Loidl, P. Przyslupski, and B. Dabrowski, "Negative refraction in ferromagnet-superconductor superlattices," Phys Rev Lett **95** (24), 247009 (2005).

35  B. Wood and J. B. Pendry, "Metamaterials at zero frequency," J Phys-Condens Mat **19** (7), 076208 (2007).

36  F. Magnus, B. Wood, J. Moore, K. Morrison, G. Perkins, J. Fyson, M. C. K. Wiltshire, D. Caplin, L. F. Cohen, and J. B. Pendry, "A d.c. magnetic metamaterial," Nat Mater **7** (4), 295-297 (2008).

37  Carles Navau, Du-Xing Chen, Alvaro Sanchez, and Nuria Del-Valle, "Magnetic properties of a dc metamaterial consisting of parallel square superconducting thin plates," Appl Phys Lett **94** (24), 242501 (2009).

38  A. H. Silver and J. E. Zimmerman, "Quantum States and Transitions in Weakly Connected Superconducting Rings," Phys Rev **157**, 317 (1967).

39  B. Chesca M. Muck, Y. Zhang, *Radio Frequency SQUIDs and Their Applications*. (Kluwer Academic Publishers, Dordrecht, 2001).

40  C. G. Du, H. Y. Chen, and S. Q. Li, "Quantum left-handed metamaterial from superconducting quantum-interference devices," Phys Rev B **74** (11), 113105 (2006).

41  N. Lazarides and G. P. Tsironis, "rf superconducting quantum interference device metamaterials," Appl Phys Lett **90** (16), 163501 (2007).

42  A. L. Rakhmanov, A. M. Zagoskin, S. Savel'ev, and F. Nori, "Quantum metamaterials: Electromagnetic waves in a Josephson qubit line," Phys Rev B **77** (14), 144507 (2008).

43  M. A. Castellanos-Beltrana and K. W. Lehnert, "Widely tunable parametric amplifier based on a superconducting quantum interference device array resonator," Appl Phys Lett **91** (8), 083509 (2007).

44  M. A. Castellanos-Beltran, K. D. Irwin, G. C. Hilton, L. R. Vale, and K. W. Lehnert, "Amplification and squeezing of quantum noise with a tunable Josephson metamaterial," Nat Phys **4** (12), 928-931 (2008).

45  M. C. K. Wiltshire, J. B. Pendry, I. R. Young, D. J. Larkman, D. J. Gilderdale, and J. V. Hajnal, "Microstructured magnetic materials for RF flux guides in magnetic resonance imaging," Science **291** (5505), 849-851 (2001).

46  S. Massaoudi and I. Huynen, "Multiple resonances in arrays of spiral resonators designed for magnetic resonance imaging," Microw Opt Techn Let **50** (7), 1945-1950 (2008).

47  C. Kurter.; S. M. Anlage, "Superconductivity takes the stage in the field of metamaterials," SPIE Newsroom **10.1117/2.1201002.002543** (2010).

48  S. M. Anlage, "High Temperature Superconducting Radio Frequency Coils for NMR Spectroscopy and Magnetic Resonance Imaging", in *Microwave Superconductivity*, edited by H. Weinstock and M. Nisenoff (Kluwer Academic Publishers, Dordrecht, 2001), pp. 337-352.

49  M. C. K. Wiltshire, J. B. Pendry, and J. V. Hajnal, "Sub-wavelength imaging at radio frequency," J Phys-Condens Mat **18** (22), L315-L321 (2006).

50  N. Engheta, "An idea for thin subwavelength cavity resonators using metamaterials with negative permittivity and permeability," IEEE Antennas Wireless Propag. Lett. **1** (1), 10-13 (2002).

51  J. D. Joannopoulos, *Photonic crystals : molding the flow of light*. (Princeton University Press, Princeton, 2008), 2nd ed.

52  C. H. Raymond Ooi, T. C. Au Yeung, C. H. Kam, and T. K. Lim, "Photonic band gap in a superconductor-dielectric superlattice," Phys Rev B **61**, 5920 (2000).

53  Yan-bin Chen, Chao Zhang, Yong-yuan Zhu, Shi-ning Zhu, and Nai-ben Ming, "Tunable photonic crystals with superconductor constituents," Mater Lett **55** (1-2), 12-16 (2002).

54  H. Takeda and K. Yoshino, "Tunable photonic band schemes in two-dimensional photonic crystals composed of copper oxide high-temperature superconductors," Phys Rev B **67**, 245109 (2003).

55    O. L. Berman, Y. E. Lozovik, S. L. Eiderman, and R. D. Coalson, "Superconducting photonic crystals: Numerical calculations of the band structure," Phys Rev B **74** (9), 092505 (2006).

56    Hesam Zandi, Alireza Kokabi, Aliakbar Jafarpour, Sina Khorasani, Mehdi Fardmanesh, and Ali Adibi, "Photonic band structure of isotropic and anisotropic Abrikosov lattices in superconductors," Physica C **467** (1/2), 8p (2007).

57    A. N. Poddubny, E. L. Lvchenko, and Y. E. Lozovik, "Low-frequency spectroscopy of superconducting photonic crystals," Solid State Commun **146** (3-4), 143-147 (2008).

58    C. Cheng, C. Xu, T. Zhou, X. F. Zhang, and Y. Xu, "Temperature dependent complex photonic band structures in two-dimensional photonic crystals composed of high-temperature superconductors," J Phys-Condens Mat **20** (27), 275203 (2008).

59    A. Aly, "Metallic and Superconducting Photonic Crystal," J Supercond Nov Magn **21** (7), 421-425 (2008).

60    H. Rauh and Y. A. Genenko, "The effect of a superconducting surface layer on the optical properties of a dielectric photonic composite," Journal of Physics: Condensed Matter **20** (14), 145203 (2008).

61    A. H. Aly, S. W. Ryu, H. T. Hsu, and C. J. Wu, "THz transmittance in one-dimensional superconducting nanomaterial-dielectric superlattice," Mater Chem Phys **113** (1), 382-384 (2009).

62    T.-H. Pei and Y.-T. Huang, "A temperature modulation photonic crystal Mach-Zehnder interferometer composed of copper oxide high-temperature superconductor," J Appl Phys **101** (8), 084502 (2007).

63    M. B. Romanowsky and F. Capasso, "Orientation-dependent Casimir force arising from highly anisotropic crystals: Application to Bi2 Sr2 Ca Cu2 O8+ delta," Phys Rev A **78**, 042110 (2008).

64    M. Tachiki, T. Koyama, and S. Takahashi, "Electromagnetic Phenomena Related to a Low-Frequency Plasma in Cuprate Superconductors," Phys Rev B **50** (10), 7065-7084 (1994).

65    D. van der Marel, "Optical Spectroscopy of plasmons and excitons in cuprate superconductors," J Supercond **17** (5), 559-575 (2004).

66    L. Feng, X. P. Liu, J. Ren, Y. F. Tang, Y. B. Chen, Y. F. Chen, and Y. Y. Zhu, "Tunable negative refractions in two-dimensional photonic crystals with superconductor constituents," J Appl Phys **97** (7), 073104 (2005).

67    D. A. Bonn, P. Dosanjh, R. Liang, and W. N. Hardy, "Evidence for Rapid Suppression of Quasi-Particle Scattering Below $T_c$ in YBa$_2$Cu$_3$O$_{7-\delta}$," Phys Rev Lett **68** (15), 2390-2393 (1992).

68    H. Takeda, K. Yoshino, and A. A. Zakhidov, "Properties of Abrikosov lattices as photonic crystals," Phys Rev B **70**, 085109 (2004).

69    Alireza Kokabi, Hesam Zandi, Sina Khorasani, and Mehdi Fardmanesh, "Precision photonic band structure calculation of Abrikosov periodic lattice in type-II superconductors," Physica C: Superconductivity **460-462** (Part 2), 1222-1223 (2007).

70    H. Zandi, A. Kokabi, A. Jafarpour, S. Khorasani, M. Fardmanesh, and A. Adibi, "Photonic band structure of isotropic and anisotropic Abrikosov lattices in superconductors," Physica C: Superconductivity **467** (1-2), 51-58 (2007).

71    Y. Wang and M. J. Lancaster, "Building left-handed transmission lines using high-temperature superconductors," J Supercond Nov Magn **20** (1), 21-24 (2007).

72    C. Caloz; T. Itoh, *Electromagnetic Metamaterials: Transmission Line Theory and Microwave Applications*. (Wiley, New York, 2005).

73    R. Y. Chiao and P. T. Parrish, "Operation of the SUPARAMP at 33 GHz," J Appl Phys **47** (6), 2639-2644 (1976).

74    E. Narimanov, private communication, 2010.

75    O. Buisson, P. Xavier, and J. Richard, "Observation of Propagating Plasma Modes in a Thin Superconducting Film," Phys Rev Lett **73**, 3153 (1994).